\documentstyle[prb,aps,epsf,twocolumn]{revtex}

\begin{document}
\draft
\title{Finite temperature spectral-functions of strongly correlated \\ 
one-dimensional electron systems}
\author{Karlo Penc\cite{*} and Mohammed Serhan}
\address{
    Max-Planck-Institut f\"ur Physik komplexer Systeme,
          Bayreuther Strasse 40, 01187 Dresden, Germany. 
}

\date{April 28, 1997}
\maketitle

\begin{abstract}
The spectral functions of $tJ$ and $tJ_{XY}$ 
models in the limit of $J/t\rightarrow 0$ and at finite temperatures 
$T\ll t$ are calculated using the spin-charge factorized wave function.
We find that the Luttinger-liquid like scaling behavior for a finite 
system with $L$ sites is restricted below temperatures of the order 
$T\lesssim J/L$. We also observe weight redistribution in
the photoemission spectral function in the energy range 
$t$, which is much larger than the temperature.\\
~
\end{abstract}


\narrowtext
 
 Single-particle spectral functions are very useful to understand 
the electronic structure
 of solids. They are measured in photoemission 
[$B(k,\omega)$] and inverse photoemission [$A(k,\omega)$] 
experiments. For actual calculations, the Lehmann representation is very 
useful:
\begin{equation}
 B(k,\omega) = \frac{1}{Z}
 \sum_{i,f,\sigma} 
  \left| \langle f | 
     a^{\phantom{\dagger}}_{k,\sigma} |i \rangle \right|^2 
    \delta(\omega - E_i + E_f) e^{-\beta E_i} ,
  \label{eq:bko}
\end{equation}
where $i$ and $f$ denote the initial and 
final states with $N$ and $N-1$ electrons, respectively, and
$a_{k,\sigma}$ annihilates an electron with momentum $k$ and spin $\sigma$.
Furthermore, $Z=\sum_i e^{-\beta E_i}$ is the partition function with 
$\beta=1/T$ being the inverse temperature. 
A similar expression holds for $A(k,\omega)$, which we will not treat in 
this paper.

In contrast to quasiparticles in usual three-dimensional Fermi liquids, 
the collective excitations of one-dimensional interacting 
electrons\cite{solyom} give rise to anomalous scaling behavior of 
the one-particle Green's function\cite{Greensfunc} with 
non-universal exponents.  For example, the momentum distribution function
$n_{k} = \frac{1}{2}\int d\omega B_\sigma(k,\omega)$ takes
the form $n_{k} \approx n_{F} + {\rm sgn}(k-k_F) |k-k_F|^\alpha$ 
near the Fermi
momentum $k_F$ and zero temperature, where the exponent $\alpha$ depends
on the actual model and coupling constants. Similarly,
the local spectral function (single-particle density of states) 
$B(\omega) = \frac{1}{L} \sum_{k} B(k,\omega)$ also scales with
a power law:\cite{voitrev}
 $B(\omega)\propto |\omega-\varepsilon_F|^\alpha$. To describe this
critical behavior of one-dimensional models 
at low energies, Haldane introduced the fruitful concept of Luttinger 
liquids.\cite{haldane}
Following a different approach, conformal field theory tells us that the 
exponents are related to the finite size corrections of the 
energy.\cite{schulz,frahm} 

 Recent experiments on  
quasi one dimensional materials raised the question if this behavior can be 
observed.\cite{exp,exp2} Furthermore, in these experiments an 
anomalous spectral weight
transfer has been observed: changing the temperature by 100 K, one can observe
weight redistribution on the scale of 1 eV, which is a hundred times larger 
than the temperature. In this paper we will try to explain this behavior 
in a simple way. 

 We are considering the isotropic and anisotropic $tJ$ model, defined by the
Hamiltonian 
\begin{eqnarray}
H_{tJ} &=& -
  t \sum_{i, \sigma} 
    (\tilde a_{i,\sigma}^\dagger  
      \tilde a_{i+1,\sigma}^{\vphantom{\dagger}} + {\rm H.c.})
    \nonumber\\
  && +  
    \sum_{i} \sum_{\alpha=x,y,z}
       J^\alpha \left( S_i^\alpha S_{i+1}^\alpha 
     - \case{1}{4} \delta_{\alpha,z} n_i n_{i+1}\right) ,
   \nonumber
\end{eqnarray}
in the limit of small exchange $J^\alpha \rightarrow 0$, 
where $\tilde a_{i,\sigma}$ are 
the usual projected operators to exclude double occupancy. Actually, the 
Hubbard model in the large-$U$ limit can be mapped onto a strong coupling 
model usually identified as the $tJ$ model plus three-site terms using a 
canonical transformation,\cite{harris} where  $J=4 t^2/U$ is small. 
The spectral function 
of the Hubbard model has been studied using exact diagonalization\cite{ED} and
 Quantum Monte Carlo \cite{montecarlo} techniques, which both have
well known limitations. 

An alternative, powerful but model limited approach is 
based on the special
property of the wave functions of the Hubbard model in the limit of large
Coulomb repulsion\cite{woyna82} (also for $J/t\rightarrow 0$ in the $tJ$
model), that the wave function factorizes:
\begin{equation}
  | i \rangle = |\psi' \rangle \otimes |\chi' \rangle .
  \label{eq:fwf}
\end{equation}
This has allowed the calculation and 
confirmation of the power law behavior of the static correlation 
function $n_k$ and gave\cite{shiba} $\alpha=1/8$.  
$|\psi' \rangle$ describes the charge degrees of freedom and 
is a wave function of free spinless fermions with momenta $k'_j$, 
quantized as $L k'_j = 2\pi {\cal I}'_j + Q'$, where ${\cal I}'_j$ 
are distinct integers, $0\leq{\cal I}'_j\leq L-1$ and $j=1,2,\dots,N$. 
Twisted boundary 
conditions are imposed by the momentum $Q'$ of the spin wave function 
$|\chi'\rangle$, which describes the 
spins on a squeezed lattice of
$N$ sites and are eigenfunctions of a spin Hamiltonian 
with an
effective spin exchange $\tilde J$ which depends on the actual charge wave
function $|\psi'\rangle$, and e.g. for the ground state 
$\tilde J^\alpha = J^\alpha n [1-\sin^2(\pi n)/(\pi n)^2]$, where $n=N/L$. 
We will take periodic boundary conditions to avoid edge effects\cite{ann} and
an even number of electrons not a multiple of four
(i.e. $N=2,6,10,\dots$) for convenience. 

To calculate the thermal average, we need to know all the energies and wave
functions of the spin model. Since for the Heisenberg
model this is very difficult to obtain, we turn to the $XY$ model 
(i.e. $J^z=0$). 
In this special case, the spin model
can be mapped onto noninteracting spinless fermions using the Wigner-Jordan
transformation. Assuming that the occupied sites represent the $\uparrow$
spins, the states are characterized by $N_\uparrow$ integer
numbers $0\ge {\cal J}'_j \ge N-1$, and the momenta $q'_j$ of 
the free spinless fermions representing the spins are quantized as 
$L q'_j = 2\pi{\cal J}'_j$.  Finally, the momentum of the spin wave function
determining the boundary condition of the charge part is 
$Q'=\sum_{j=1}^{N_\uparrow} q'_j=2\pi{\cal J}'/N$, with ${\cal J}'$
integer. 
The energy of the state is simply $E_i = E_{i,c} + E_{i,s}$, where 
$E_{i,c} = -2 t \sum_{j=1}^N \cos k'_j$ and 
$E_{i,s} = \tilde J_{XY} \sum_{j=1}^{N_\uparrow} \cos q'_j$,
while the momentum reads $P_i = \sum_{j=1}^N k'_j$. 
One should note that despite the fact
that both the charge and spin wave functions in Eq.~(\ref{eq:fwf}) are those 
of free spinless fermions, the resulting wave function describes a nontrivial
and strongly correlated system. As far as the exponent $\alpha$ 
(at $T = 0$) is concerned,
it changes from $\alpha=1/8$ in the isotropic case to $\alpha=1/4$ in the $XY$ 
case.\cite{XYtJ}

Similarly, the final, $N-1$ electron wave function factorizes as well: 
$| f \rangle = |\psi \rangle \otimes |\chi \rangle$. 
The quantum numbers for the spinless fermions
representing the charges are ${\cal I}_j$, and the corresponding momenta
$L k_j = 2\pi {\cal I}_j + Q$. Here $Q=2\pi {\cal J}/(N-1)$ is the 
momentum of the $N-1$ spin wave function, ($0\ge {\cal J} \ge N-2$).

Since the charge and the spin part are coupled through the momentum $Q'$ of
the spin wave function, the partition function 
does not factorize\cite{hatsugai} (i.e. the free energy is not a sum of 
charge and spin contribution) 
and it will read $Z= \sum_{Q'} Z_c(Q') Z_s(Q')$,
where 
\[
  Z_c(Q') = \sum_{\{{\cal I}'_j\}} e^{-\beta E_{i,c}},
 \quad 
  Z_s(Q') = \sum_{\{{\cal J}'_j\}_{Q'}} e^{-\beta E_{i,s}},
\] 
and the sum in $Z_s$ is over the states with given momentum $Q'$.
In calculating the thermodynamic averages, 
one can work in principle 
in an ensemble fixing either the magnetization or the magnetic 
field. We have used both ensembles, and although the results in the
thermodynamic limit should be independent of the ensemble we choose, 
there are strong finite size effects.

Even though we know all the excitations for the $tJ_{XY}$ model, 
we will make further restrictions
which are needed to perform calculations on reasonably large system sizes: 
Namely, we will consider temperatures much smaller than the energy scale of
the charges. In other words, for the charge part we neglect the excitations 
and take the ground state given by consecutive 
integers $\{{\cal I}'\}=\{-N/2,-N/2+1,\dots,N/2-1\}$. 
Then the remaining free parameter is  
$T/\tilde J$, and all the temperature dependence is now in the spin part.
Furthermore, since the energy of the charge part also depends 
on $Q'$ as $E_{i,c}(Q')-E_{i,c}(Q'=\pi) = \frac{1}{2 \pi L} u_c (Q'-\pi)^2$,
where $u_c \propto t$ is the charge velocity,
we will assume that the momentum of the spin part in the initial $N$
electron state is $Q'=\pi$. This
restriction is actually more for convenience, as the result does not depend
on this assumption - we will comment on this later on.

 Using the factorized wave function, the spectral function defined in 
Eq.~(\ref{eq:bko}) simplifies to\cite{local}
\begin{equation}
  B(k,\omega) 
  =
 \sum_{Q,\sigma} 
  D_{\sigma}(Q,\beta)
  B_{Q}(k,\omega) .
  \label{eq:blhbca} 
\end{equation}
Here $B_{Q}(k,\omega) = B_{Q,Q'=\pi}(k,\omega)$ depends on the spinless 
fermion wave function only:
\begin{eqnarray}
  B_{Q,Q'}(k,\omega) 
 & = & L
 \sum_{\{I\}} 
  \left| 
    \langle \psi_{Q} |
    b^{\phantom{\dagger}}_{0} 
    | \psi'_{Q'} \rangle
    \right|^2\nonumber\\
   &&\times 
  \delta(\omega - E_{i,c} + E_{f,c} ) 
  \delta_{k, P_{i} \!-\! P_{f}} ,
 \nonumber
\end{eqnarray}
where $b_0$ annihilates a spinless fermion at site $0$. The matrix 
elements in $B_{Q,Q'}(k,\omega)$ read:
\begin{eqnarray}
  &&L  \left| 
    \langle \psi_{Q} |
    b^{\phantom{\dagger}}_{0} 
    | \psi'_{Q'} \rangle
    \right|^2
  = L^{-2N+2} 
  \sin^{2N-2}\frac{Q' \!-\! Q}{2}
  \nonumber\\
  &&\times 
   \prod_{j>i} \sin^2 \frac{k_j \!-\! k_i}{2} 
   \prod_{j>i} \sin^2 \frac{k'_j \!-\! k'_i}{2} 
   \prod_{i,j} \sin^{-2} \frac{k'_i \!-\! k_j}{2}  .
 \nonumber
\end{eqnarray}
We can actually recognize Anderson's orthogonality catastrophe\cite{anderson} 
in  these complicated matrix elements, which is a consequence of 
changing the boundary condition from $Q$ to $Q'$ in the charge wave 
function due to momentum transferred to the spins.

 On the other hand, the contribution 
of the spin degrees of freedom\cite{sorella2,local}
$D_{\sigma}(Q,\beta)=D_{\sigma}(Q,Q'=\pi,\beta)$ 
is given by
\[
  D_\sigma(Q,Q',\beta) = 
  \frac{1}{N-1} \frac{1}{Z_s}
  \sum_{m,\{{\cal J'}\}_{Q'}} \omega_{0\rightarrow m,\sigma} 
                 e^{im(Q'-Q)} e^{-\beta E_{i,s}},
\]
where $\omega_{0\rightarrow m,\sigma}$ denotes the amplitude to 
transfer a $\sigma$ spin from site $0$ to $m$:
\begin{equation}
  \omega_{0\rightarrow m,\sigma} = 
  \langle \chi' | 
    \hat P_{m,m-1}\dots \hat P_{1,0} n_0 
  | \chi' \rangle .
   \label{eq:wdef}
\end{equation} 
The operator $\hat P_{j,j+1}=2 {\bf S}_j {\bf S}_{j+1} +\frac{1}{2}$ 
permutes the spins on sites $j$ and $j+1$.

\paragraph{Spin part:} 

For the $XY$ model, after introducing the spinless fermions 
(with operators $f$) in the Wigner-Jordan transformation, 
the permutation operator reads
\[
 \hat P_{j+1,j} = n_{j+1} n_{j} + f^\dagger_{j+1} f^{\phantom{\dagger}}_{j} 
             + (1 \!-\! n_{j+1})(1 \!-\! n_{j}) + 
               f^\dagger_{j} f^{\phantom{\dagger}}_{j+1} ,
\]
and the spin transfer amplitude can be easily calculated from 
Eq.~(\ref{eq:wdef}) using Wick's theorem. We find that 
\[
  \omega_{0\rightarrow m,\uparrow} = 
   \left|
     \begin{array}{cccc}
       g_0 & g_1 & \dots & g_{m} \\
      1 \!+\! g_{-1} & g_0 & \dots & g_{m-1}\\
      1 \!+\! g_{-2} & 1 \!+\! g_{-1} & \dots & g_{m-2}\\
      \vdots & \vdots & & \vdots \\
      1 \!+\! g_{-m} & 1 \!+\! g_{1-m}& \dots& g_0 \\
     \end{array}
   \right| ,
\]
 where 
$ g_{l} = (-1)^l 
  \langle \chi | f^\dagger_{l} f^{\phantom{\dagger}}_{0}  | \chi\rangle 
  =  \frac{1}{N} \sum_{j=1}^{N_\uparrow} e^{i (\pi-q'_j) l}
$. In particular, $g_0 = N_\uparrow/N$ and $g_{-j} = g_{j}^*$, furthermore
the relation 
$ \omega_{0\rightarrow N-1-m,\sigma} = e^{i Q'}  \omega^*_{0\rightarrow
m,\sigma}$ holds.

  For large temperatures $t\gg T \gg \tilde J$ (equivalent to ``hot spins''
of Ref.~\onlinecite{gebhard})
a high temperature expansion is possible:
if we relax the constraint that we take only states with momenta $Q'$, then
it follows that $Z_s= 2^N + O(\beta J_{XY})$ and for 
$\omega_{0\rightarrow m,\sigma}$ we have to count the number of states where
the first $m+1$ spins have $S^z=\uparrow$, which is $2^{N-m-1}$. 
 Working in a subspace with definite momentum $Q'$,
each subspace will acquire roughly $1/N$ of the values given above (the
actual distribution depends on how many states are in a given $Q'$ subspace),
 and in the thermodynamic limit we get
\begin{equation}
  D_\sigma(Q,Q',\beta\rightarrow 0) = 
   \frac{1}{N-1} \frac{3}{10-8 \cos(Q-Q')} .
  \label{eq:hte}
\end{equation}
This result is not only valid for the $XY$ model, but also for the 
isotropic Heisenberg model.

\begin{figure}
\epsfxsize=8.5 truecm
\centerline{\epsffile{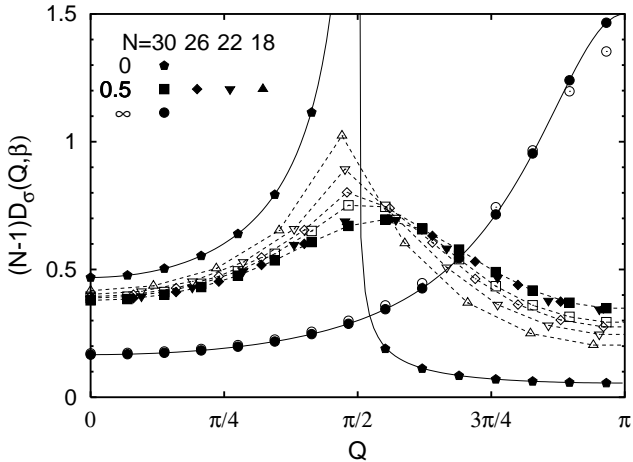}}
\caption{
Temperature dependence of $D_\sigma(Q,\beta)$ for the 
$XY$ model in zero magnetic field
(solid) and zero magnetization (empty symbols) for 
$T/\tilde J=0$, 0.5 and $T\gg \tilde J$. 
The solid line for $T=0$ shows the $N=250$ result, and for  
$T\gg \tilde J$  Eq.~(\protect{\ref{eq:hte}}) is plotted.
}
\label{fig:DQ}
\end{figure}

  We show the behavior of the spin part in Fig.~\ref{fig:DQ}. Apart from
the clear power-law singularity near $Q=\pi/2$ at zero temperature,
we observe that at fixed small temperature this behavior disappears as we
increase the system size. This indicates that the singularity will 
vanish for any finite temperature in the thermodynamic limit.
Also there is a difference between calculating $D_\sigma(Q,\beta)$ 
in the two 
ensembles mentioned above, however the finite size
effects are decreasing with increasing $N$.   
Let us also note that the sum rule 
$\sum_Q D_\sigma(Q,\beta) = N_\sigma/N$ is satisfied for any temperature.
Furthermore, $D_\sigma(Q,Q',\beta) \approx D_\sigma(Q-Q'+\pi,\beta)$ in the 
thermodynamic limit.

\paragraph{Momentum distribution function:}

 From Eq.~(\ref{eq:blhbca}) we get:
\begin{equation}
  n_{k} =  \sum_Q B_Q(k) D_{\sigma}(Q,\beta) .
  \label{eq:nkf}
\end{equation}
   To calculate $n_{k}$ efficiently, we have to find a 
convenient way to evaluate $B_Q(k)$. For that reason, let us follow
Ref.~\onlinecite{shiba}: In the alternative representation of the
 momentum distribution 
\[
 n_k = \frac{1}{Z}\sum_i \sum_{l=0}^{L-1}
      \langle i | 
        a^\dagger_{l,\sigma} a^{\phantom{\dagger}}_{0,\sigma} 
      | i \rangle e^{i k l}  e^{-\beta E_i}  
\]
we replace $| i \rangle$ by the factorized wave function, 
Eq.~(\ref{eq:fwf}):
\begin{equation}
   \langle i | 
      a^\dagger_{l,\sigma} a^{\phantom{\dagger}}_{0,\sigma} 
   | i \rangle 
  = \sum_{m=0}^{N-2}
  \langle \psi' | c^\dagger_{l} 
          \delta_{N_{l}-m} c^{\phantom{\dagger}}_{0} 
  | \psi' \rangle
  \omega_{0\rightarrow m,\sigma} ,
  \label{eq:cjcjp}
\end{equation}
where $N_{l}=\sum_{l'=0}^{l} n_{l'}$ counts the number of spinless 
fermions between sites $0$ and $l$,  and $\omega_{0\rightarrow m,\sigma}$ 
is calculated for the particular $|\chi'\rangle$.
Now, replacing $\delta_{N_{l}-m}$ by its Fourier representation:
\[ \delta_{N_{l}-m} = \frac{1}{N-1} \sum_{Q} e^{i (Q-Q')(N_{l}-m)},
\] 
and comparing Eqs.~(\ref{eq:nkf}) and (\ref{eq:cjcjp}), we get 
\begin{equation}
 B_{Q,Q'}(k) 
  = \sum_{l} e^{ikl} 
   \langle \psi' | c^\dagger_{l}
    \prod_{ l'=1}^{l-1} e^{i n_{l'}(Q-Q')}  
    c^{\phantom{\dagger}}_{0} 
    | \psi' \rangle .
   \label{eq:BQkalt}
\end{equation} 
This can be further simplified using the identity
$e^{i n_l (Q-Q')} = 1 + x n_l $, where we introduced $x=e^{i(Q-Q')}-1$,
so that 
$ \langle  \psi' | c^\dagger_{l}
    \prod_{ l'=1}^{l-1} e^{i n_{l'}(Q-Q')}  
    c^{\phantom{\dagger}}_{0} 
  | \psi' \rangle 
$ in Eq.~(\ref{eq:BQkalt}) is equal to
\[
  \left|
   \begin{array}{cccc}
       \langle c^\dagger_l c^{\phantom{\dagger}}_0\rangle &
       \langle c^\dagger_l c^{\phantom{\dagger}}_1\rangle &
       \dots &
       \langle c^\dagger_l c^{\phantom{\dagger}}_{l-1} \rangle
     \\
         x \langle c^\dagger_1 c^{\phantom{\dagger}}_0\rangle &
   1\!+\!x \langle c^\dagger_1 c^{\phantom{\dagger}}_1\rangle &
       \dots &
         x \langle c^\dagger_1 c^{\phantom{\dagger}}_{l-1}\rangle 
     \\  
         x \langle c^\dagger_2 c^{\phantom{\dagger}}_0\rangle &
         x \langle c^\dagger_2 c^{\phantom{\dagger}}_1\rangle &
       \dots &
         x \langle c^\dagger_1 c^{\phantom{\dagger}}_{l-1}\rangle 
     \\  
      \vdots & \vdots & & \vdots \\
     \\
         x \langle c^\dagger_{l-1} c^{\phantom{\dagger}}_0\rangle &
         x \langle c^\dagger_{l-1} c^{\phantom{\dagger}}_1\rangle &
       \dots &
    1\!+\!x \langle c^\dagger_{l-1} c^{\phantom{\dagger}}_{l-1}\rangle 
   \end{array}
  \right|  ,
\]
where $\langle c^\dagger_{l} c^{\phantom{\dagger}}_{l'}\rangle = 
  \frac{1}{L}\sum_j e^{-ik'_j (l-l')}$. Using this equation, we are able to
compute $B_Q(k,\omega)$ for systems with a few hundred sites. It turns out
  that $B_{Q,Q'}(k,\omega) = B_{Q-Q'+\pi}(k,\omega)$ apart from some small
  finite size corrections, therefore our assumption to fix $Q'=\pi$ is 
justified. 

\begin{figure}
\epsfxsize=8.5 truecm
\centerline{\epsffile{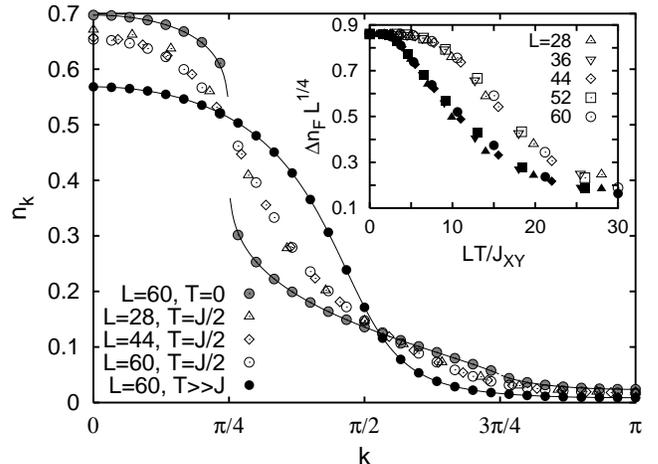}}
\caption{
 Momentum distribution of the $tJ_{XY}$ model for 
$T=0$ (solid line: $L=500$), $T/\tilde J = 0.5$,
 and $t \gg T \gg \tilde J$ (solid line: $L=300$)
for quarter filling ($L=2N$) and zero magnetic field.
There are strong finite size effects for $T=\tilde J/2$. In the insert
we show the scaling of $\Delta n_F$ for zero magnetic field (solid) and
zero magnetization (empty symbols) obtained from different system sizes and 
temperatures. 
}
\label{fig:nk}
\end{figure}

We show our numerical results in Fig.~\ref{fig:nk}.  The $T=0$
result shows power law behavior at the Fermi momentum. 
Increasing the temperature, the power
law survives until some crossover temperature ($\approx \tilde J/L$), 
where it becomes a continuous function of momentum for higher 
temperatures. To study 
this behavior in  detail, we concentrate on the jump at $k_F$, 
defined as $\Delta n_F = n_{k^-_F}-n_{k^+_F}$, where 
$k^\pm_F = k_F \pm \pi  N/L$ are the momenta of the finite system 
closest to the Fermi point. This jump  
is finite for finite size system and scales with $L^{-\alpha}$ in the
  Luttinger-liquid. If the singularity disappears and $n_k$ becomes a
  continuous function around $k_F$, then $\Delta n_F \propto 1/L$.
In the inset of Fig.~\ref{fig:nk} we show the ``size independent'' 
$L^{\alpha}\Delta n_F$ vs. $LT/\tilde J$. It is remarkable, that at low
  temperature the points follow a universal curve:
\[
  \Delta n_F = L^{-\alpha} f(L T/\tilde J).
\]
 A crossover temperature, scaling with $\tilde J/L$,
can be clearly observed, and for larger temperatures $L^{\alpha} \Delta n_F
\rightarrow 0$.
 This behavior can be understood if we recall that the temperature enters by 
dividing the energy $\propto 2\pi u_\sigma/L $ of the  low-energy 
excitations. 

\begin{figure}
\epsfxsize=8.5 truecm
\centerline{\epsffile{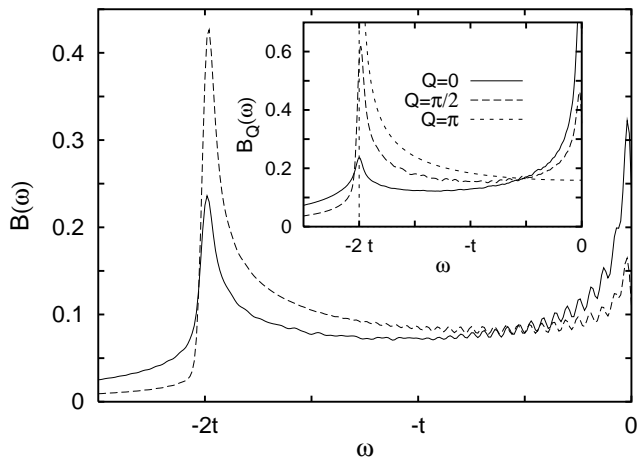}}
\caption{
 Local spectral function for $T=0$ (solid) and  $t \gg T \gg \tilde J$ (dashed
 line) for the quarter-filled Hubbard model, $L=220$. 
For this particular filling $\varepsilon_F=0$.
In the inset: $B_Q(\omega)$ for different values of $Q$. 
}
\label{fig:bom}
\end{figure}

\paragraph{Local spectral function:} 
The single-particle density of states is given by
\begin{equation}
  B(\omega) =   \sum_{Q,\sigma} D_{\sigma}(Q,\beta)
                B_{Q}(\omega) ,
  \label{eq:alocal}
\end{equation}
where $B_{Q}(\omega) = \frac{1}{L} \sum_{k} B_{Q}(k,\omega)$. Let us
concentrate on the isotropic $tJ$ model (equivalent to 
large-$U$ Hubbard model) 
in the limiting $T=0$ and $t\gg T\gg \tilde J$ cases only.
 At low temperatures $D_\sigma(Q,\beta)$ is large near $Q=\pi/2$,
and the largest part in the convolution 
(\ref{eq:alocal}) comes from $B_{Q=\pi/2}(\omega)$. For the ``hot spin'' case,
$D_\sigma(Q,\beta)$ is 
large near $Q=\pi$, and $B_{Q=\pi}(\omega)$ gives most of
the contribution to $B(\omega)$, shown in Fig.~\ref{fig:bom}. In other words, 
increasing the temperature we transfer less and less momentum to the spins,
 and the role of the orthogonality catastrophe in $B_Q(\omega)$ decreases.  
Since changing $Q$ results in a considerable redistribution of
 the weight in $B_{Q}(\omega)$ (see the inset in
Fig.~\ref{fig:bom}), the weight transfer of $B(\omega)$ at the energy scale 
of $t$ is due to the temperature dependence of $D_\sigma(Q,\beta)$ 
set on a much smaller
temperature scale -- naively we would expect smearing of $B(\omega)$ 
near the Fermi energy within $|\omega-\varepsilon_F| \approx T$. 
We should also note that the divergence of the spectral
function at the Fermi energy is purely the artifact of the $J/t\rightarrow 0$
 limit\cite{local}. 
For finite $J$, the local spectral function has a broad peak around 
$\omega \approx \frac{\pi}{2}\tilde J$ 
due to the spinon dispersion, and a second
broad peak near the band edge ($\omega \approx 2\tilde t$). 
The weight transfer then would be from the``spinon''  to the 
``holon'' peak. A similar weight redistribution is observed in the
two-dimensional $tJ$ model as well.\cite{jaklic}

To conclude, we have studied the temperature evolution of the momentum 
distribution function and local spectral function. First, we give a 
method to calculate $n_k$ for large system sizes for the $tJ_{XY}$ model
at zero temperature. Next, we observed that the power-law behavior is
restricted to temperatures inversely proportional to the system size. In the
thermodynamic limit the system is critical at $T=0$ only. Finally, a
weight redistribution in the single-particle density of states takes 
place over a broad energy range, which can be 
easily understood using the concept of ``spin-charge'' separation.

 We would like to thank H.~Frahm, J. Jakli\v c, H.~Shiba and W. Stephan for 
stimulating discussions.

\end{document}